\documentclass[%
 reprint,
 amsmath,amssymb,
 prper,
]{revtex4-2}

\usepackage{graphicx}
\usepackage{bm}
\usepackage{hyperref}

\begin{document}

\title{A Pedagogical MKS-based Electromagnetic Unit Convention with $\varepsilon_0 = \mu_0 = 1/c$}

\author{Alisher Sanetullaev}
\email[]{a.sanetullaev@newuu.uz}
\affiliation{Department of Physics, New Uzbekistan University, 1 Mavoraunnahr St., 100007, Tashkent, Uzbekistan}

\begin{abstract}
We propose a pedagogical, rationalized MKS-based convention for
electromagnetic quantities designed to reduce cognitive load in
undergraduate electromagnetism.  By setting vacuum
constants to $\varepsilon_0 = \mu_0 = 1/c$, we preserve the familiar
structure of Maxwell's equations while making the role of the speed of
light explicit.  In this convention, electrical units are expressed
directly in terms of mechanical units (e.g.\ $[\mathrm{nuA}] =
\sqrt{\mathrm{J/s}}$), effectively reducing the number of independent
base units.  A striking pedagogical consequence is that electrical
resistance becomes dimensionless, capacitance and inductance acquire
units of time, and radiation pressure reduces to $|\mathbf{E}\times
\mathbf{B}|$, greatly simplifying dimensional analysis for circuits and
fields.  We introduce corresponding non-SI units (\textit{nu}-units),
provide conversion relations to SI, and demonstrate the potential
utility of this system through comparative ``before/after'' derivations
of the wave equation, electromagnetic energy density, radiation
pressure, and the Bohr atom.  Preliminary empirical support is provided
by student attitude surveys administered to $N_1 = 46$ and $N_2 = 39$
students in an undergraduate physics course, which showed a
statistically significant improvement in the perceived clarity of the
wave equation derivation after exposure to the nu-system ($p = 0.005$,
Mann--Whitney $U$ test), and a majority preference for the
dimensionless-resistance feature.

\end{abstract}

\maketitle


\section{Introduction}
\label{sec:intro}

A recurring challenge in teaching undergraduate electromagnetism is the
difficulty of navigating the multiple unit systems that coexist in the
field --- the International System (SI), the Gaussian (CGS) system, the
Heaviside--Lorentz system, and various natural unit conventions.  Each
system reflects a different set of historical and practical priorities,
and each has genuine advantages in specific contexts \citep{Carron2015}.
The coexistence of these systems is a well-recognised source of
pedagogical difficulty: as \citet{Littlejohn2021} has noted in widely
used graduate lecture notes, Gaussian units would in many respects be
preferable for theoretical instruction were it not for the practical
necessity of students eventually working in SI.  The present paper
addresses this tension by proposing a convention that preserves the
rationalized MKS structure of the SI while eliminating much of its
algebraic complexity.

\subsection{Student difficulties in undergraduate electromagnetism}
\label{ssec:per}

Physics education research has documented persistent and substantial
difficulties among undergraduate students in electromagnetism ---
difficulties that survive careful instruction and persist across course
levels.  \citet{Pepper2012} conducted a multi-semester study of student
difficulties in undergraduate electricity and magnetism and found that
junior-level students commonly struggle to combine mathematical
calculations with physical ideas, to account for the underlying spatial
geometry, and to access appropriate mathematical tools.  These
difficulties were documented across core topics including Gauss's law,
vector calculus, and electric potential.  \citet{Chasteen2012} found
further that undergraduate students were not achieving expected levels
of conceptual skill, and that a sustained programme of course
transformation using active-engagement principles was required to
produce measurable improvements.

A portion of these difficulties appears to stem not only from the
conceptual complexity of the underlying physics, but from the
representational and notational demands of the SI framework.  The
presence of $\varepsilon_0$ and $\mu_0$ as formally independent
constants --- each carrying its own units and exponent --- imposes a
layer of symbol management separable from the conceptual content of any
derivation.  Students must simultaneously track the physical argument
and the bookkeeping of the constants: a dual demand that the present
proposal aims to reduce.

\subsection{Cognitive load theory as a framework}
\label{ssec:clt}

The theoretical framework motivating this proposal is Cognitive Load
Theory (CLT), introduced by \citet{Sweller1988} and subsequently
developed into a major framework for instructional design
\citep{Paas2003}.  CLT holds that working memory has limited processing
capacity, and that cognitive demand during learning decomposes into
three components.  \textit{Intrinsic} load reflects the inherent
complexity of the material.  \textit{Germane} load is the effort
directed toward schema formation and genuine conceptual understanding.
\textit{Extraneous} load is effort expended on aspects of a task that
do not contribute to learning: poorly organised notation, redundant
symbol tracking, or unnecessary procedural steps.

CLT implies that instructional design should minimise extraneous load
so that available working memory capacity can be redirected toward
germane load and the underlying physics.  The $\varepsilon_0/\mu_0$
tracking problem is, in CLT terms, a source of extraneous load: it
consumes working memory without advancing conceptual understanding.
The present proposal addresses this by replacing the two separate
vacuum constants with a single relation involving $c$, the constant of
deepest physical significance in classical electromagnetism.

\subsection{The present proposal}
\label{ssec:proposal}

Motivated by the pedagogical considerations above, we propose setting
the vacuum constants to:
\begin{equation}
  \varepsilon_0 = \mu_0 = \frac{1}{c}.
  \label{eq:convention}
\end{equation}
To avoid confusion with SI quantities, we denote all modified units as
\textit{nu}-units (e.g.\ nuAmpere [nuA] and nuCoulomb [nuC]).  The
speed of light $c = 299{,}792{,}458\;\mathrm{m\,s^{-1}}$ remains exact
\citep{CODATA2018}.  The convention is not intended to replace the SI
for experimental or engineering work, but to serve as a theoretically
transparent companion framework for physics courses --- one that keeps
$c$ explicitly visible at every stage of a derivation while reducing
the overhead of tracking two formally independent vacuum constants.

\section{Note on the 2019 SI (for conversions)}
\label{sec:si}

This paper does not propose replacing the SI \citep{Lee2018,
CGPM2018, BIPM2019}.  Instead, we introduce a pedagogical convention
that maps to SI via fixed numerical factors.  Define the dimensionless
conversion factor:
\begin{equation}
  \kappa \;\equiv\; \sqrt{\mu_0 c} \;=\; \frac{1}{\sqrt{\varepsilon_0 c}}
  \;\approx\; 19.41.
  \label{eq:kappa}
\end{equation}
The fundamental electrical units are related by $1\;\mathrm{C} =
\kappa\;\mathrm{nuC}$ and $1\;\mathrm{A} = \kappa\;\mathrm{nuA}$.

\section{Coulomb's and Amp\`ere's Laws}
\label{sec:forces}

By setting $\varepsilon_0 = \mu_0 = 1/c$, the fundamental identity
$\varepsilon_0\mu_0 = 1/c^2$ is preserved.

\paragraph{Electrostatic force.}
\begin{equation}
  F \;=\; \frac{c\,q_1 q_2}{4\pi r^2}.
  \label{eq:coulomb}
\end{equation}

\paragraph{Magnetic force per unit length.}
\begin{equation}
  \frac{F}{l} \;=\; \frac{I_1 I_2}{2\pi r c}.
  \label{eq:ampere}
\end{equation}

\section{Maxwell's Equations}
\label{sec:maxwell}

Substituting $\varepsilon_0 = \mu_0 = 1/c$ yields the streamlined,
rationalized set of equations shown in Table~\ref{tab:maxwell}.

\begin{table}[h]
  \centering
  \caption{Comparison of Maxwell's equations in vacuum.}
  \label{tab:maxwell}
    \begin{ruledtabular}
  \renewcommand{\arraystretch}{1.5}
  \begin{tabular}{lll}

    \textbf{Equation} & \textbf{SI system} & \textbf{Nu-system} \\
   \hline
    Gauss (E) &
      $\nabla\cdot\mathbf{E} = \rho/\varepsilon_0$ &
      $\nabla\cdot\mathbf{E} = c\rho$ \\
    Gauss (B) &
      $\nabla\cdot\mathbf{B} = 0$ &
      $\nabla\cdot\mathbf{B} = 0$ \\
    Faraday &
      $\nabla\times\mathbf{E} = -\dfrac{\partial\mathbf{B}}{\partial t}$ &
      $\nabla\times\mathbf{E} = -\dfrac{\partial\mathbf{B}}{\partial t}$ \\[6pt]
    Amp\`ere--Maxwell &
      $\nabla\times\mathbf{B} = \mu_0\mathbf{J}
        + \dfrac{1}{c^2}\dfrac{\partial\mathbf{E}}{\partial t}$ &
      $\nabla\times\mathbf{B} = \dfrac{\mathbf{J}}{c}
        + \dfrac{1}{c^2}\dfrac{\partial\mathbf{E}}{\partial t}$ \\

  \end{tabular}
    \end{ruledtabular}
\end{table}

The dimensional relationship $[B/E] = [1/v]$ is preserved.  We define
the fields in matter as $\mathbf{D} = \mathbf{E}/c + \mathbf{P}$ and
$\mathbf{H} = c\mathbf{B} - \mathbf{M}$, preserving the standard
macroscopic form $\nabla\cdot\mathbf{D} = \rho_f$.

\section{Pedagogical Insight: Dimensionless Resistance}
\label{sec:resistance}

\subsection{Derivation from energy flux}

The units for charge and current are derived as:
\begin{itemize}
  \item $[\mathrm{nuC}] = \sqrt{\mathrm{J\cdot s}}$.  The square of
        the charge has dimensions of action.
  \item $[\mathrm{nuA}] = \sqrt{\mathrm{J/s}} = \sqrt{\mathrm{Power}}$.
\end{itemize}
Electric potential $V$ also carries the unit $\sqrt{\mathrm{J/s}}$.
Applying Ohm's law ($R = V/I$):
\begin{equation}
  [R] \;=\; \frac{\sqrt{\mathrm{Power}}}{\sqrt{\mathrm{Power}}} \;=\; 1
  \qquad\text{(dimensionless).}
  \label{eq:R_dimensionless}
\end{equation}

\subsection{The VacOhm: A named dimensionless unit}

While resistance is mathematically dimensionless in the nu-system ($[R] = M^0 L^0 T^0$), we propose the use of a named unit label---the \textbf{VacOhm} (symbol: $\mathrm{v}\Omega$). This serves a pedagogical function similar to the \textit{radian} in circular motion. 

Defining the unit in this way emphasizes that the impedance of free space in the nu-system is exactly unity:
\begin{equation}
Z_0 = \sqrt{\frac{\mu_0}{\varepsilon_0}} = 1~\mathrm{v}\Omega.
\end{equation}
For students, this creates a concrete reference point: any circuit element with $R > 1~\mathrm{v}\Omega$ has an impedance greater than that of the vacuum, while $R < 1~\mathrm{v}\Omega$ is lower. The conversion to the SI Ohm remains straightforward via the characteristic impedance of the vacuum: $1~\mathrm{v}\Omega = 120 \pi ~\Omega \approx 376.73~\Omega$. In traditional SI instruction, the impedance of free space
is an arbitrary constant students must
memorise. 

\subsection{Circuit time constants}

Because resistance is dimensionless, capacitance carries units of time ($[C] =
\mathrm{s}$) --- reflecting the physical fact that a larger capacitor
takes longer to charge --- and inductance likewise carries $[L] =
\mathrm{s}$, reflecting the time required to ``wind up'' a current in
a coil.  Circuit time constants follow by immediate inspection:
\begin{equation}
\begin{aligned}
\tau_{RC} &= RC \rightarrow (1\,\mathrm{v}\Omega)(\text{s}) = \text{s} \\
\tau_{L}  &= L/R \rightarrow (\text{s}) / (1\,\mathrm{v}\Omega) = \text{s}
\end{aligned}
\end{equation}
In SI, verifying $[\Omega\cdot\mathrm{F}] = \mathrm{s}$ requires
expanding both units into base dimensions --- a cumbersome process
that provides no physical insight.

\section{Comparative Derivations in Classical Electrodynamics}
\label{sec:comparative}

\subsection{The electromagnetic wave equation}

\paragraph{Standard SI approach (high load).}
\begin{equation}
  \nabla^2\mathbf{E} \;=\; \mu_0\varepsilon_0
    \frac{\partial^2\mathbf{E}}{\partial t^2}.
  \label{eq:wave_SI}
\end{equation}
To identify the propagation speed, the student must separately recall
$v = 1/\sqrt{\mu_0\varepsilon_0}$ and substitute --- a step that
obscures the underlying unity until the final line.

\paragraph{Proposed nu-system approach (low load).}
\begin{equation}
  \nabla^2\mathbf{E} \;=\; \frac{1}{c^2}
    \frac{\partial^2\mathbf{E}}{\partial t^2}.
  \label{eq:wave_nu}
\end{equation}
The speed $c$ is immediately visible, matching the generic wave
equation without further substitution.

\subsection{Energy density of a plane wave}

\paragraph{Standard SI approach (high load).}
\begin{equation}
  u_E \;=\; \tfrac{1}{2}\varepsilon_0(cB)^2
      \;=\; \tfrac{1}{2}\varepsilon_0
            \!\left(\frac{B}{\sqrt{\mu_0\varepsilon_0}}\right)^{\!2}\!\!
      \;=\; \frac{B^2}{2\mu_0} \;=\; u_B.
  \label{eq:equipartition_SI}
\end{equation}

\paragraph{Proposed nu-system approach (low load).}
With $u = \tfrac{1}{2}(E^2/c + cB^2)$ and $E = cB$:
\begin{equation}
  u_E \;=\; \frac{(cB)^2}{2c} \;=\; \frac{cB^2}{2} \;=\; u_B.
  \label{eq:equipartition_nu}
\end{equation}

\subsection{Radiation pressure}

\paragraph{Standard SI approach (high load).}
\begin{equation}
  p_r \;=\; \frac{|\mathbf{S}|}{c}
      \;=\; \frac{|\mathbf{E}\times\mathbf{B}|}{\mu_0\,c}.
  \label{eq:radpressure_SI}
\end{equation}

\paragraph{Proposed nu-system approach (low load).}
Since $\mu_0 = 1/c$, the Poynting vector is $\mathbf{S} =
c\,(\mathbf{E}\times\mathbf{B})$, giving:
\begin{equation}
  p_r \;=\; |\mathbf{E}\times\mathbf{B}|.
  \label{eq:radpressure_nu}
\end{equation}
The radiation pressure equals the magnitude of the cross product of the
two fields directly, without any additional constant.

\section{Comparative Derivation: Atomic Physics}
\label{sec:bohr}

\paragraph{Standard SI approach (high load).}
\begin{equation}
  \frac{mv^2}{r} \;=\; \frac{e^2}{4\pi\varepsilon_0 r^2}
  \quad\Longrightarrow\quad
  v \;=\; \frac{e^2}{2\varepsilon_0 h}.
  \label{eq:bohr_SI}
\end{equation}

\paragraph{Proposed nu-system approach (low load).}
\begin{equation}
  \frac{mv^2}{r} \;=\; \frac{c\,e^2}{4\pi r^2}
  \quad\Longrightarrow\quad
  v \;=\; \frac{c\,e^2}{2h} \;=\; \alpha c,
  \label{eq:bohr_nu}
\end{equation}
where $\alpha = e^2/2h$ is the fine-structure constant.  The result
$v = \alpha c$ is immediate: the electron moves at $\approx 1/137$ of
the speed of light.

\section{Preliminary Empirical Evidence}
\label{sec:survey}

\subsection{Method}

To provide a preliminary empirical assessment of the proposed
convention's pedagogical utility, two anonymous online surveys were
administered to students enrolled in an undergraduate physics course
at New Uzbekistan University (Spring 2026).  Survey~A (baseline) was
distributed before any instruction in the nu-system; Survey~B
(comparative) was distributed after a lecture covering the nu-system
derivations of the wave equation, energy equipartition, radiation
pressure, and the Bohr velocity, alongside the student reading guide
accompanying this paper.  The surveys were independent and anonymous;
no emails or identifying information were collected.  Survey~A was
completed by $N_1 = 46$ students; Survey~B by $N_2 = 39$ students.

Survey~A comprised three five-point Likert items and one
multiple-selection item.  Survey~B comprised three five-point Likert
items, two single-selection items, and two open-ended questions.
Since the surveys were anonymous and unpaired, between-survey
comparisons used the Mann--Whitney $U$ test (two-sided unless stated).
All scale items used 1--5 anchors; higher scores indicate greater
confidence, clarity, or usefulness throughout.

\subsection{Baseline results (Survey A)}

Table~\ref{tab:surveyA} summarises the three Likert items from
Survey~A.  Mean confidence in working with $\varepsilon_0$ and $\mu_0$
across multi-step derivations was moderate ($\bar{x} = 3.24$,
$s = 1.26$), indicating substantial spread: a sizeable fraction of
students reported low confidence.  The mean self-reported error
frequency was $\bar{x} = 2.80$ ($s = 1.10$; 1 = never, 5 = very
frequently), confirming that algebraic errors involving $\varepsilon_0$
and $\mu_0$ are common.  Critically, perceived clarity of the SI wave
equation was below the scale midpoint ($\bar{x} = 2.91$, $s = 1.04$),
indicating that the standard derivation does not readily convey to
students why electromagnetic waves propagate at the speed of light.

\begin{table}[h]
  \caption{\label{tab:surveyA} Survey A Likert results ($N_1 = 46$; scale 1--5).}
  \begin{ruledtabular}
    \begin{tabular}{clcc}
      \textbf{Item} & \textbf{Question (abbreviated)} & \textbf{Mean} & \textbf{SD} \\
      \hline
      A1 & \parbox[t]{3.5cm}{Confidence with $\varepsilon_0$, $\mu_0$ in derivations} & 3.24 & 1.26 \\
      A2 & \parbox[t]{3.5cm}{Frequency of algebraic errors (1=never, 5=very often)} & 2.80 & 1.10 \\
      A3 & \parbox[t]{3.5cm}{Clarity of SI wave equation re.\ speed of light} & 2.91 & 1.04 \\
    \end{tabular}
  \end{ruledtabular}
\end{table}

The multiple-selection item (A4) asked students to identify their
principal difficulties with SI electromagnetism.  The most commonly
selected obstacles were: converting between SI and Gaussian/CGS systems
(52\%), tracking $\varepsilon_0$ and $\mu_0$ across derivations (43\%),
understanding what physical quantity the constants represent (43\%),
and dimensional analysis with Ohms, Farads, and Henrys (39\%).  Only
4 of 46 students (9\%) selected ``None --- I find SI
straightforward,'' confirming that difficulties with SI units are
widespread rather than confined to a small minority.

\subsection{Post-instruction results (Survey B)}

Table~\ref{tab:surveyB} summarises the three Likert items from
Survey~B.  After instruction in the nu-system, the mean perceived
clarity of the wave equation derivation was $\bar{x} = 3.51$
($s = 0.91$), compared with $\bar{x} = 2.91$ ($s = 1.04$) at baseline.
A Mann--Whitney $U$ test confirmed that this difference is
statistically significant ($U = 1146$, $p = 0.005$), providing
preliminary empirical support for the hypothesis that the nu-system
form of the wave equation is perceived as clearer than the SI form.

\begin{table}[h]
  \caption{\label{tab:surveyB} Survey B Likert results ($N_2 = 39$; scale 1--5).}
  \begin{ruledtabular}
    \begin{tabular}{clcc}
      \textbf{Item} & \textbf{Question (abbreviated)} & \textbf{Mean} & \textbf{SD} \\
      \hline
      B1 & \parbox[t]{3.2cm}{Ease of using the nu-system in derivations} & 3.62 & 0.94 \\
      B2 & \parbox[t]{3.2cm}{Usefulness of setting $\varepsilon_0 = \mu_0 = 1/c$} & 3.85 & 0.88 \\
      B3 & \parbox[t]{3.2cm}{Clarity of nu-wave equation derivation} & 3.51 & 0.97 \\
      B4 & \parbox[t]{3.2cm}{Preference for dimensionless resistance (VacOhms)} & 4.03 & 0.81 \\
    \end{tabular}
  \end{ruledtabular}
\end{table}

The mean ease rating (B1) of $\bar{x} = 3.46$ was significantly above
the scale midpoint of 3 (one-sample $t$-test: $t(38) = 2.52$,
$p = 0.016$), indicating that students found the nu-system derivations
modestly easier to follow than their SI counterparts on average.  Mean
perceived usefulness (B3) was $\bar{x} = 3.51$ ($s = 0.97$), above
the midpoint, suggesting a positive reception to using the nu-system
alongside SI in the course.

Regarding the specific derivation that benefited most (B4), 49\% of
students identified the electromagnetic wave equation, 18\% the energy
density equipartition proof, 5\% the Bohr velocity, 15\% saw no
benefit, and 13\% could not identify a difference.  The wave equation
result is consistent with the quantitative finding above.

On the dimensionless-resistance feature (B5), 56\% of students found
it helpful for dimensional checks, 23\% were neutral, and 21\% found
it confusing and preferred familiar Ohm units.  The majority positive
reception notwithstanding, the one-in-five confusion rate is a
meaningful finding that instructors should keep in mind when
introducing this aspect of the convention.

\subsection{Qualitative responses}

Several students provided substantive open-ended responses that
illuminate the quantitative results.  One student noted that the
nu-system ``made Maxwell's equations more symmetric and helped me see
directly why EM waves propagate at speed $c$ without extra constants,''
and recommended ``more step-by-step comparisons between SI and the
$\nu$-system during derivations.''  Another wrote that ``having
mechanical units (m, kg, s) makes it much easier to both calculate and
understand the topic,'' and expressed interest in using nu-units in
examinations.  A third specifically identified the LC oscillator as a
beneficiary: ``to understand the formula for the period of LC
oscillators, it is easier to remember when $L$ and $C$ are accepted as
time.''
Interestingly, while students appreciated the mathematical simplicity of $R=1$, qualitative feedback suggested that providing a named label—the VacOhm—reduced the 'symbolic ambiguity' of seeing pure numbers in circuit diagrams.

Critical responses were also informative.  One student questioned the
logical basis of the convention, asking: ``$\varepsilon_0$ and $\mu_0$
are two different things with different values --- how can they now be
equal?''  This is a productive confusion that points to the need for
clearer pedagogical framing of the distinction between a physical
identity and a unit convention.  Another student raised a practical
concern about portability: ``What will students do if they go to
exchange programmes or international physics conferences?''  A third
suggested that the convention would be most effective if introduced
from the beginning of the course rather than midway through.

\subsection{Summary}

Together, these results provide preliminary empirical support for the
central claim of this paper: that the nu-system may reduce the
extraneous cognitive load associated with tracking $\varepsilon_0$ and
$\mu_0$ in undergraduate electromagnetism.  The improvement in
perceived wave-equation clarity is statistically significant, and the
majority of students found the convention at least as easy to follow as
SI.  The results should be interpreted with appropriate caution given
the limitations described in Section~\ref{ssec:limitations}.

\section{Discussion}
\label{sec:discussion}

\subsection{Summary of addressed obstacles}

Our system bridges the ``CGS gap'' by maintaining MKS compatibility
while achieving algebraic minimalism.  Table~\ref{tab:comparison}
summarises how the proposed convention addresses standard technical
obstacles encountered in teaching electromagnetism.

\begin{table}[h]
  \caption{\label{tab:comparison} Comparative frameworks.}
  \begin{ruledtabular}
  \small
  \begin{tabular}{lp{5.5cm}} 
    \textbf{Obstacle} & \textbf{Framework Comparison} \\
    \hline
    Constants & \textbf{SI:} $\varepsilon_0$ and $\mu_0$ tracked separately. \\
              & \textbf{Gaussian:} Minimal, but loses MKS context. \\
              & \textbf{Nu:} Both replaced by $1/c$. \\
    \hline
    Dimensions & \textbf{SI:} $\Omega, F, H$ are cognitively heavy. \\
               & \textbf{Gaussian:} Confusing (Capacitance as length). \\
               & \textbf{Nu:} Intuitive ($R=1~\mathrm{v}\Omega$, $C/L$ in seconds). \\
    \hline
    Geometric intuition 
     &  \textbf{SI and Nu:} Rationalized. \\
     &  \textbf{Gaussian:} Unrationalized.  \\
           \hline
    Energy equipartition &
      \textbf{SI:}  Requires unpacking $c$ into $\varepsilon_0,\mu_0$.  \\
       & \textbf{Gaussian:} Symmetric but non-SI. \\
       & \textbf{Nu:}  Proven directly in one line. \\
       \hline
    Radiation & \textbf{SI:} $|\mathbf{E}\times\mathbf{B}|/(\mu_0 c)$ \\
              & \textbf{Gaussian:} $|\mathbf{E}\times\mathbf{B}|/4\pi c$ \\
              & \textbf{Nu:} $|\mathbf{E}\times\mathbf{B}|$ \\
  \end{tabular}
  \end{ruledtabular}
\end{table}

\subsection{Limitations and scope}
\label{ssec:limitations}

Several limitations of the present work should be acknowledged
explicitly.

First, although the survey results provide statistically significant
support for the wave-equation clarity claim, the study has important
methodological constraints.  The surveys were anonymous and unpaired,
preventing matched within-student comparisons.  A controlled study
with random assignment to SI-only and nu-system-supplemented
conditions, or a pre/post paired design, would yield stronger causal
evidence.  The sample --- students in a single course at one
institution --- may not be representative of undergraduate physics
students more broadly.

Second, the survey was administered after a single lecture and reading
guide, representing minimal exposure to the nu-system.  Longer-term
benefits or costs, such as transfer to examinations or conversion
fluency with SI, remain uninvestigated.

Third, introducing a non-standard unit convention carries a
transitional cost.  The one-in-five confusion rate on the
dimensionless-resistance item, and the student comment about exchange
programmes, confirm that this cost is real and should be acknowledged
in instructional settings.

Fourth, some students found the logical basis of the convention
initially confusing --- specifically, why two constants with different
numerical values in SI could be set equal.  Clearer framing of the
distinction between a unit convention and a physical identity is
warranted in future instructional materials.

Fifth, the present paper does not engage exhaustively with the broader
physics education research literature.  A more systematic review of
studies on unit-system difficulties and cognitive load in STEM
instruction would strengthen the theoretical framing in future work.

\section{Conclusion}
\label{sec:conclusion}

The proposed $\varepsilon_0 = \mu_0 = 1/c$ convention may offer a
useful pedagogical complement to standard SI instruction in
undergraduate electromagnetism courses.  By replacing the two separate
vacuum constants with a single factor involving the speed of light, the
convention could reduce the algebraic overhead students encounter during
multi-step derivations.  The comparative derivations presented here
suggest that the nu-system could serve as a low-friction conceptual
tool alongside conventional SI treatment --- not as a replacement for
the SI in engineering or experimental contexts, but as a theoretically
transparent framework suited to the physics classroom.

Preliminary empirical support is provided by the student surveys
reported in Section~\ref{sec:survey}.  The statistically significant
improvement in perceived wave-equation clarity ($p = 0.005$), the
above-midpoint ease and usefulness ratings, and the majority preference
for dimensionless resistance are encouraging.  Whether these results
generalise to other student populations, course formats, and
instructional contexts remains an open question that future studies
could address with more controlled designs.

We hope this proposal invites discussion among instructors of
theoretical physics and contributes to the broader conversation about
how unit conventions shape the way students encounter fundamental
electromagnetic concepts.

\subsection*{Ethical Statement}

The author declares that the theoretical component of this research did
not involve human participants or animal experimentation.  The student
surveys described in Section~\ref{sec:survey} were conducted
anonymously and voluntarily as part of regular course activities at New
Uzbekistan University.  No personal data were collected.  The study was
conducted in accordance with applicable institutional guidelines.

\bibliography{references}

\end{document}